\documentclass[twocolumn,english,prb,showpacs]{revtex4}
\usepackage[latin9]{inputenc}
\usepackage{amssymb}
\usepackage{graphicx}
\usepackage{amsmath}
\usepackage{color}
\usepackage{mathrsfs}
\usepackage{float}
\usepackage{indentfirst}
\usepackage{mathrsfs}	
\usepackage{float}
\usepackage{indentfirst}
\usepackage{textcomp}

\newcommand\ccite[1]    {~\cite{#1}}
\newcommand\PLOTFILE[1]  {./#1}

\newcommand{\COMMENTED}[1]{}

\begin{document}
\title{Symmetry in Auxiliary-Field Quantum Monte Carlo Calculations }

\author{Hao Shi}
\author{Shiwei Zhang}

\affiliation{Department of Physics,
             The College of William and Mary,
             Williamsburg, Virginia 23187}

\begin{abstract}
We show how symmetry properties can be used to greatly increase the accuracy and
efficiency in auxiliary-field quantum Monte Carlo (AFQMC) calculations of electronic systems. With the Hubbard model
as an example, we study symmetry preservation in two aspects of ground-state AFQMC calculations, 
the Hubbard-Stratonovich transformation and the form of the trial wave function. 
It is shown that significant improvement over state-of-the-art calculations can be achieved. 
In unconstrained calculations, the implementation of symmetry often leads to shorter convergence time 
and much smaller statistical errors, thereby a substantial reduction of the sign problem.
Moreover, certain excited states become possible to calculate which are otherwise beyond reach. In
calculations with constraints, the use of symmetry can reduce the systematic
error from the constraint. It also allows release-constraint calculations, leading to essentially exact results in many cases. 
Detailed comparisons are made with exact diagonalization results.
Accurate ground-state energies are then presented for larger system sizes in the
two-dimensional repulsive Hubbard model.
\end{abstract}

\pacs{71.10.Fd, 02.70.Ss, 05.30.Fk}

\maketitle

\section{Introduction}\label{sec:Introduction}

The study of interacting quantum many-body
systems remains an outstanding challenge, especially systems with strong particle
interactions, where perturbative approaches are ineffective.
Numerical simulations provide a 
promising approach for studying such systems.
One of the most general numerical approaches is quantum Monte Carlo (QMC) methods based on 
auxiliary-fields, which are applied 
 in condensed matter physics, nuclear physics,
high-energy physics, and quantum chemistry.
These methods
\cite{BSS1PhysRevD.24.2278,BSS2PhysRevB.24.4295,KooninSugiyama19861} allow essentially exact calculations of ground-state
and finite-temperature equilibrium properties of interacting many
fermion systems. As is well known, however, they suffer from the sign problem 
which severely limits their applicability \cite{SIGNPPRB1990, PhysRevB.55.7464}. A key 
issue in algorithm development in this context is 
to reduce the convergence time or variance reduction for fixed computing cost.

Considerable progress has been achieved in circumventing this problem by constraining the 
random walks in sampling the space of auxiliary-fields. These methods have come under the name
of Constrained Path Monte Carlo (CPMC) 
\cite{PhysRevB.55.7464,PhysRevB.78.165101}
for systems where there is a sign problem (for 
example, Hubbard-like models where the local interactions lead to auxiliary-fields that are real).
For electronic systems where there is a 
phase problem (as the  
Coulomb interaction leads to complex fields), 
the methods \cite{PhysRevLett.90.136401,PhysRevE.70.056702, al-saidi:224101} have been referred to  as 
phaseless or phase-free auxiliary-field QMC.
In both cases, 
the idea is to constrain the sign or phase of the overlap of the sampled Slater determinants with a trial wave function. 
It eliminates the sign or phase instability and restores low-power (typically 
to the third power of system size) computational scaling. Applications
to a variety of systems have shown that the methods are very accurate, 
even with simple trial wave wave functions taken directly from mean-field calculations
(see, e.g. Refs~\cite{PhysRevB.80.214116,PhysRevLett.104.116402} and references therein). 
However, these methods are approximate. For example, open-shell situations often result in 
larger systematic errors. It is thus important to understand and develop ways to improve the 
quality of the constraint.

Symmetry properties and projection have a long history in quantum many body problems (see, e.g. Refs~\cite{nuclearsym,ImadaPRB2004,PhysRevB.72.224518,PhysRevB.72.085116,PhysRevE.77.026705,JPSJ.77.114701,PhysRevB.85.245130})
In this paper, we discuss the imposition of symmetry properties and their effects in AFQMC calculations. 
It is shown that symmetry can often be rigorously preserved in the AFQMC framework 
despite the stochastic nature of the calculations, and taking advantage of this can lead to 
dramatic improvements.
We address the issue from two aspects: symmetry preserved projection and 
the symmetry of the trial wave function.  

In the symmetry preserved projection, we study the form of the Hubbard-Stratonovich (HS)
transformation and its effect on the computational efficiency and accuracy. We focus on 
the comparison between charge-decompositions which preserve SU(2) symmetry
and the standard form of Hirsch spin-decomposition. 
We discuss the optimal form of the charge decomposition which involves subtracting the proper ``mean-field" background \emph{prior to} making the HS transformation.

With a trial wave function which imposes exact symmetry properties, one is able to reduce the projection time as
well as the statistical error bar.  
Symmetry alleviates the sign problem, since the convergence time to the ground state 
is reduced by ensuring the right symmetry in the trial wave function used to initiate the projection.
 Furthermore, the correct symmetry 
can help sustain projections of excited states which have different symmetry from that 
of the ground state. The most important feature of preserving symmetry in the trial wave function is
in constrained AFQMC calculations, where we will show that the quality of the approximation
improves significantly.

We consider the effect of symmetry in three different forms of ground-state AFQMC calculations, 
unconstrained Free Projection (FP), constrained path (CP) calculations, 
and released-constraint (RC) calculations. 
The first, FP, corresponds to the standard algorithm which is formally exact but has the sign problem. The third, RC, can be thought of in the present context
as a more sophisticated version of FP, in which we start from a much better initial state, namely a converged CP run, and 
release (i.e., remove) the constraint and continue the projection. In FP and RC, the focus is on 
the use of symmetry to reduce the statistical fluctuations and prolong 
the imaginary-time projections. We show that improvements are seen in both methods.
Especially in RC the use of symmetry can lead to a dramatic effects and significantly 
ameliorate the sign problem.

We are interested in studying the effect of symmetry in CP calculations, in order to 
further improve the constraints.
We find that the use of trial wave functions
which preserve symmetry can greatly reduce the systematic bias from the constraint. 
In open-shell systems, the use of simple trial wave functions which restore symmetry properties
eliminates most of the bias from the constraint. Another key finding is that one
can switch between different 
forms of the HS transformation in CP and the following RC calculations. This is motivated by considering
the different behaviors of statistical errors and sensitivity to the constraint. 

We will use the two-dimensional repulsive Hubbard model as an example in our discussions.
The model provides  an ideal test ground for the present study. On the one-hand, it is computationally simple  
where exact diagonalization results are available in non-trivial system sizes.
On the other hand, it is a fundamental model 
which captures many of the key aspects of simulating correlated materials and
which, in fact, contains many unanswered questions.
Through the approach presented in this paper, significantly more accurate calculations 
can be done on larger systems. For example,
essentially exact results can now be obtained for 
$> 100$ sites, as shown below.

The remainder of the paper is organized as follows. In section \ref{sec:AFQMC}, we 
summarize the relevant features of the AFQMC method in three different forms: 
the unconstrained FP,  CP, and RC, in order to 
 facilitate the ensuing discussions.
In section \ref{sec:symmetry}, we describe separately the two aspects of symmetry
improved AFQMC, namely symmetry-preserving HS transformation and 
symmetry in the trial wave function. Results are first shown for demonstration and 
benchmark, then followed by new, near-exact ground-state energies for larger system sizes.
Finally we conclude in Sec.~\ref{sec:conoutl}.

\section{Auxiliary Field Quantum Monte Carlo Methods} \label{sec:AFQMC}
 We first summarize the key features of  ground state AFQMC methods that are relevant to 
 the symmetry study to follow.
The three different forms of AFQMC are summarized in three subsections below.
As mentioned, we will use the two-dimensional repulsive Hubbard model as a concrete example; however most of our discussions will apply to other Hamiltonians, including more realistic materials Hamiltonians.

The Hubbard model is written in second-quantized form as \ccite{J.hubbard.1963}:
\begin{equation}
 \hat{H}=\hat K +\hat V=-t\sum_{\langle i,j \rangle \sigma}^{L} c_{i\sigma}^{\dagger}c_{j\sigma}+U\sum_{i}^{L} n_{i\uparrow}n_{i\downarrow}\,.
 \label{eq:HubHamil}
\end{equation}
Here $L$ is the number of lattice sites, $c_{i\sigma}^{\dagger}$ and $c_{i\sigma}$ are creation and annihilation operators of an electron
of spin $\sigma$ on the $i$-th lattice site, $t$ is the nearest-neighbor hopping energy and $U$ is the interaction strength.
 Throughout this paper, we will use $t$ as units of energy and set $t=1$.
We assume that there are $N_{\uparrow}$ spin-up electrons and $N_{\downarrow}$ spin-down electrons on the lattice.

The Hamiltonian in Eq.~(\ref{eq:HubHamil}), whose Hilbert space grows
exponentially in size with $L$, presents an enormous challenge. Questions remain open about its properties.
Ground state projection with Monte Carlo (MC) sampling is one of the candidates to solve the problem.
The projection method is:
 \begin{equation}
 |\psi_{0}\rangle \propto \lim_{\beta\rightarrow\infty} e^{-\beta (\hat{H}-E_{T})} |\psi_{T}\rangle\,,
\end{equation}
where $E_{T}$ and
$|\psi_{T}\rangle$ 
are guesses of the ground state energy and wave function, 
and $\langle \psi_{0}|\psi_{T}\rangle\neq0$ 
in order for the projection to yield the ground state asymptotically. To target a lowest energy excited state of a different symmetry from $|\psi_{0}\rangle$ is similar to doing a ground-state calculation, except
one would choose a $|\psi_{T}\rangle$ which is not orthogonal to the targeted excited state but satisfies $\langle \psi_{0}|\psi_{T}\rangle=0$.

The propagator can be evaluated using a Trotter-Suzuki breakup \ccite{trotter1959,MasuoSuzuki1976}:
 \begin{equation}
 (e^{-\varDelta \tau(\hat{K}+\hat{V})})^n=(e^{-\frac{1}{2}\varDelta \tau\hat{K}}e^{-\varDelta \tau\hat{V}}
e^{-\frac{1}{2}\varDelta \tau\hat{K}})^n+O(\varDelta \tau^2)\,.
\end{equation}
Here we have $\varDelta \tau n=\beta$, and a Trotter error arises from the omission of the higher
order terms. 
We will not be concerned with the Trotter error here, other than to note that
it can be controlled by
extrapolating to $\varDelta \tau\rightarrow0$ with separate calculations using different $\varDelta \tau$ values. 
In the results shown in this paper, we either perform such an extrapolation explicitly, 
or have checked via separate calculations that the Trotter error is 
within the statistical error. We also mention that, for the Hubbard interaction, 
decompositions without Trotter errors are possible \ccite{PhysRevB.84.241110}.

The two-body propagator is then decoupled into one-body propagator by 
auxiliary fields, using the HS transformation \ccite{PhysRevLett.3.77,PhysRevB.28.4059}. The general form is:
\begin{equation}
e^{-\varDelta \tau\hat{V}}=\sum_x p(x) e^{\hat{o}(x)}\,,
\label{eq:HS1}
\end{equation}
where $\hat{o}(x)$ is a one-body operator that depends on the auxiliary field $x$, $p(x)$ is a probability density function with
the normalization $\sum_{x} p(x)=1$. 
In general the sum in Eq.~(\ref{eq:HS1}) is an integral, and $x$ is a many-dimensional vector
whose dimension is of the order of the size of the one-particle basis. In the Hubbard model, 
$x$ typically has $L$ components, one for each lattice site. 
By setting:
\begin{equation}
\hat{B}(x)=e^{-\frac{1}{2}\varDelta \tau\hat{K}} e^{\hat{o}(x)} e^{-\frac{1}{2}\varDelta \tau\hat{K}},
\end{equation}
we rewrite the projection as
 \begin{equation}
 |\psi_{0}\rangle= \sum_{\overrightarrow{X}} P(\overrightarrow{X}) \prod_{i=1}^n \hat{B}(x_{i})|\psi_{T}\rangle\,.
 \label{eq:projection0}
\end{equation}
The vector$\overrightarrow{X}$  means $(x_{1},x_{2},\cdots,x_{n})$, and $P(\overrightarrow{X})=\prod_{i}p(x_{i})$.

Then the ground state properties can be evaluated by:
\begin{equation}
 \langle \hat{A}\rangle_{0}=\frac{\langle\psi_{0}|\hat{A}|\psi_{0}\rangle}{\langle\psi_{0}|\psi_{0}\rangle}\,,
\end{equation}
which are many-dimensional integrals (e.g.,  $2nL$-dimensions in the Hubbard model). MC methods are used to calculate the high dimension integrals, by sampling 
the probability density function using the Metropolis algorithm\ccite{metropolis:1087} or a related method. The sign problem emerges because the integrand in the denominator, 
$P(\overrightarrow{X}) \langle \psi_{T}|\prod \hat B(x_i) | \psi_{T}
\rangle$
 is not always positive, which causes the MC signal to be eventually lost in the sampling noise.

\subsection{Unconstrained Free Projection (FP)}
\label{ssec: method_FP}

We carry out the FP calculation \cite{jacobi:43,PhysRevLett.90.136401,PhysRevB.80.214116} with an open-ended random 
walk similar to the CP approach \ccite{PhysRevB.55.7464,PhysRevLett.90.136401}. 
However, the discussion on symmetry in FP calculations in this paper will apply directly
to standard ground-state AFQMC calculations using Metropolis sampling of 
 paths.
In our approach, a population of $N_w$ random walkers is carried, which are typically initialized by the trial 
 wave function. Each walker will have a weight $w$ whose value is set as one at the beginning of the projection: 
 \begin{equation}
  |\psi^{(0)}\rangle=\sum_{i}^{N_{w}} w_{i}^{(0)} |\phi_{i}^{(0)}\rangle
\label{eqn:fpwalk}
 \end{equation}

We apply the projection in Eq.~(\ref{eq:projection0}) by random walks in Slater determinant space,
instead of using the usual approach of Metropolis sampling of auxiliary-field paths. In each step,
we sample the auxiliary field $x$ according to 
 $p(x)$ by MC, and apply $\hat{B}(x)$ to the Slater determinant wave function.
  Since the operators 
only contain one-body terms, they will generate another Slater determinant \ccite{PhysRevB.41.11352}:
 \begin{eqnarray}
  |\psi^{(1)}\rangle &=&\sum_{i}^{N_w}\sum_{x_{i}} p(x_{i})  \hat{B}(x_{i}) w_{i}^{(0)} |\phi_{i}^{(0)}\rangle \nonumber  \\
&=&\sum_{i}^{N_w} w_{i}^{(1)} |\phi_{i}^{(1)}\rangle
 \end{eqnarray}

 During the projection
we multiply the constant (non-operator) values of the formula, e.g., the overall normalization $e^{\varDelta\tau E_{T}}$, 
to the weight of the walker. The weight  of each walker will fluctuate in 
the random walk and
after a few steps, some walkers can have large weights and some walkers will have small weights. We apply 
a population control procedure, splitting the walkers with large weights and eliminating
 walkers with small
 weights with the appropriate probability \ccite{PhysRevB.57.11446}, such that 
the overall probability distribution is preserved but the weights are made more uniform. It will introduce a 
population control bias \cite{PhysRevB.55.7464}. In our calculations, the population control bias is much smaller
than the statistical error. This is easy to accomplish in CP calculations. In FP and RC calculations, large
populations are necessary as discussed below.
 Modified Gram-Schmidt orthogonalization is applied to each walker 
 periodically as well \ccite{PhysRevB.40.506}.

As mentioned, the FP calculations are done in the same framework as the 
CP calculations. 
This is different from standard Metropolis sampling \ccite{BSS1PhysRevD.24.2278,SorellaEUL,ImadaJPSJ.65.189}, 
which keeps an entire path of a 
\emph{fixed length} $n$ as the object to sample. This framework does not have any ergodicity 
issues  \ccite{PhysRevB.44.10502}, and it is straightforward to project to longer imaginary-time
in order to approach the ground state. We typically turn off importance sampling in FP,  
sampling the fields according to $P(\overrightarrow{X})$ instead of using either 
the force bias  \ccite{PhysRevLett.90.136401,PhysRevB.80.214116} 
or direct importance-sampling of discrete fields  \ccite{PhysRevB.55.7464} as is done in 
CP calculations. Empirically we find that this tends to give smaller statistical errors than 
invoking importance sampling and then lifting the constraint.
The use of population control helps 
to reduce the noise but ultimately the shortcoming of this approach is 
that the lack of importance sampling will cause large noises as system size or $n$ increases.
Since in these situations the sign problem, when uncontrolled, tends to 
overwhelm the calculation anyway, the shortcoming is not of major practical relevance, and
we find this mode of sampling to often be the more efficient in practice.

At the $n^{\rm th}$ step in the propagation, we measure the energy by:
 \begin{equation}
  E=\frac{\displaystyle\sum\limits_{i}^{N_w} w_{i}^{(n)}\langle \psi_{T} |H|\phi_{i}^{(n)}\rangle}
{\displaystyle\sum\limits_{i}^{N_w} w_{i}^{(n)}\langle \psi_{T}|\phi_{i}^{(n)}\rangle}\,.
\label{eqn:fpmear}
\end{equation}
If the projection has equilibrated, we can combine the populations at multiple $n$ values in the 
estimator above to improve statistics on the ground-state energy.
The energy measure in Eq.~(\ref{eqn:fpmear}) is variational if we set $|\psi^{(0)}\rangle=|\psi_{T}\rangle$, since
 \begin{equation}
  E=\frac{ \langle \psi_{T} |H e^{-\beta H}|\psi_{T}\rangle}
{\langle \psi_{T}| e^{-\beta H}|\psi_{T}\rangle}=
\frac{ \langle \psi_{T} |e^{-\beta H/2 }H e^{-\beta H/2 }|\psi_{T}\rangle}
{\langle \psi_{T}| e^{-\beta H/2 }e^{-\beta H/2 }|\psi_{T}\rangle}\,.
\end{equation}
To calculate the expectation value of an observable which does not commute with the Hamiltonian, 
we can use  back-propogation \ccite{PhysRevB.55.7464,PhysRevE.70.056702} using part of the 
path and projecting the trial wave function $\langle \psi_{T} |$. Because of the lack of importance sampling, back-propagation will 
tend to be very noisy in FP, and large population size will typically be needed.

The application of symmetry in either the trial wave function or the propagator $\hat B(x)$ is 
straightforward in FP.
In Sec.~\ref{sec:symmetry} we show how the use of symmetry can improve FP 
calculations and allow longer projection time with smaller statistical fluctuations. Furthermore,
some excited states can be accessed. 

\subsection{Constraint Path (CP)}

The constrained path (CP) approximation allows one to eliminate the sign problem present in FP.
During the FP steps, the overlap between the ground state and the projected wave function, 
$\langle \psi_{0}|\psi^{(l)}\rangle$, will in general approach zero, because 
intrinsically the projection is symmetric about $|\psi^{(l)}\rangle$ and $-|\psi^{(l)}\rangle$. 
In other words, at any given imaginary time $l$, the projection would proceed identically if each 
random walker $|\phi^{(l)}\rangle$ were switched to $-|\phi^{(l)}\rangle$, for example by 
a permutation of two of its orbitals. This means that, unless the random walks are somehow 
strictly confined to only one kind of ``sign'', it will invariably become a random and equal mixture of both,
given sufficiently large $l$. Thus measurements from 
the MC sampling will eventually have infinite variance.
This is how the sign problem appears in an FP calculation. 

The CP approach is based on the observation 
that \ccite{PhysRevB.55.7464}, if any particular walker has the zero overlap with the ground state 
at imaginary time $\tau_l\equiv l\Delta \tau$ in the projection:
\begin{equation}
 \langle \psi_{0}|\phi^{(l)}\rangle=0
\end{equation}
this walker will contribute zero at any future time $\beta> \tau_l$, because:
\begin{equation}
 \langle \psi_{0}|e^{-(\beta-\tau_{l})\hat{H}}|\phi^{(l)}\rangle=0\,.
\end{equation}
Then we are able to discard the walker once its path reaches a point where the overlap becomes zero. 
With this constraint the sign problem is eliminated, and the projection will
still lead to the exact ground state. However, we obviously do not know the exact ground state wave function. In CP calculations, 
a trial wave function, $|\psi_{T}\rangle$, is chosen
for determining the sign of the overlap. A walker which develops a zero overlap with $|\psi_{T}\rangle$
during the projection is discarded.

Importance sampling can be introduced  in CP calculations both as a natural way to impose the 
constraint and for variance reduction \ccite{PhysRevB.55.7464,PhysRevLett.90.136401,PhysRevE.70.056702}.
With importance sampling,  the wave function during the projection can be written as:
\begin{equation}
 |\psi^{(l)}\rangle=\sum_{i}^{N_w} w_{i}^{(l)} \frac{ |\phi_{i}^{(l)}\rangle}{\langle \psi_{T}|\phi_{i}^{(l)}\rangle}\,.
\label{eqn:cpwalk}
\end{equation}
Instead of $p(x)$, one samples the auxiliary-fields from 
\begin{equation}
 \widetilde{p}(x)=p(x) \frac{\langle \psi_{T}|\hat{B}(x)|\phi_{i}^{(l)}\rangle}{\langle \psi_{T}|\phi_{i}^{(l)}\rangle}\,,
\end{equation}
which can be accomplished either directly for discrete fields using a heat bath-like approach
\ccite{PhysRevB.55.7464} or more generally 
via a force bias  \ccite{PhysRevLett.90.136401,PhysRevE.70.056702}.
This will automatically prevent the random walks from sampling any determinants with  zero (or negative) overlap with the trial wave function. Those with larger overlap will be sampled more, although the weight from importance sampling will ensure that the exact distribution 
defined by Eq.~(\ref{eqn:cpwalk}) is sampled.
The energy can be calculated by the mixed estimate similar to Eq.~(\ref{eqn:fpmear}), 
although now with importance sampling it has the form: 
 \begin{equation}
  E=\frac{\displaystyle\sum\limits_{i}^{N_w} w_{i}^{(n)} \frac{\langle \psi_{T} |H|\phi_{i}^{(n)}\rangle}{\langle \psi_{T}|\phi_{i}^{(n)}\rangle}}
{\displaystyle\sum\limits_{i}^{N_w} w_{i}^{(n)}}\,.
\label{eqn:CPmeas}
\end{equation}
Following diffusion Monte Carlo (DMC), we refer to the quantity in the numerator, 
$E_L(\phi)\equiv \langle \psi_{T} |H|\phi\rangle/\langle \psi_{T}|\phi\rangle$ as the local
energy. An important characteristic of the constrained path approximation is that the 
mixed estimate in Eq.~(\ref{eqn:CPmeas}) is not variational \cite{PhysRevB.59.12788}.

The CP approximation has proved very accurate in the Hubbard model, especially for closed shell systems \ccite{PhysRevB.55.7464}. For instance,
the energy at $U=4$ is typically within
$< 0.5\%$ of the exact diagonalization result \cite{PhysRevB.78.165101}.
It is, however, approximate. The systematic error in the energy tends to be larger in
open-shell systems. Here we show that this error can be significantly reduced with trial wave functions which observe the correct 
symmetry.

\subsection{Release Constraint (RC)}

 From a converged CP calculation, one can release the constraint and continue with the projection.
Calculations of similar character have been done in the framework
of DMC, 
under the name of released node \ccite{ceperley:5833}.
Since the CP result is already very close to the ground state and FP in AFQMC 
tends to have a reduced sign problem in general, one can 
expect that releasing the constraint in AFQMC will be effective and will
allow the removal or reduction of systematic 
bias in more systems.  

In principle the idea of releasing the constraint is straightforward. As mentioned, the RC calculation can theoretically be viewed as an FP calculation, with a much better starting point.  
Technically, however, the implementation of RC can be challenging. 
The initial population, namely that from CP, is obtained with importance sampling, which automatically imposes the constraint.
 On the one hand,  the importance function must 
be modified in RC to allow the random walks to have a significant chance to sample the region with 
$ \langle \psi_T|\phi^{(l)}\rangle$ being negative (or develop different phases in the more 
general case). Indeed, the more efficient approach in FP in AFQMC 
seems to be with no importance function, as discussed earlier. On the other hand, since the initial 
population  from CP  has the form of Eq.~(\ref{eqn:cpwalk}),  ``undoing"
the importance sampling will cause large statistical fluctuations and accelerate the onset of the
sign problem. Thus we need to find the best way to 
interface  the RC part with CP so as
to minimize the growth of statistical noise. 

RC calculations within AFQMC have been published by Sorella \ccite{PhysRevB.84.241110} recently.  
Our approach  is somewhat different  and we will publish the 
method and further results elsewhere. The algorithmic details
do not affect the issues that are the focus of the present paper, namely the use of 
symmetry properties to improve the  calculations. A key aspect is our 
use of different forms of the HS transformation in the CP and the RC portions of the calculations.
That is, we switch to a different HS decomposition in the RC in order to impose exact symmetry 
properties which drastically change the behavior of the RC calculations, as described below.

In RC calculations, we 
will use a mixed estimator similar to Eq.~(\ref{eqn:fpmear}) to measure the energy:
 \begin{equation}
  E_{\rm RC}(\beta)=\frac{\displaystyle\sum\limits_{i}^{N_w} w_{i}^{\rm CP}\langle \psi_{T} |H e^{- \beta \hat{H}}|\phi_{i}^{\rm CP}\rangle}
{\displaystyle\sum\limits_{i}^{N_w} w_{i}^{\rm CP}\langle \psi_{T}|e^{-\beta \hat{H}}|\phi_{i}^{\rm CP}\rangle}\,.
\label{eq:E_mixed_RC}
\end{equation}
Thus $E_{\rm RC}(\beta=0)=E_{\rm CP}$.
As mentioned, the mixed estimate in CP is not variational. The RC energy will asymptotically converge to the exact ground-state energy. However, it can converge from below or above. 
Indeed, as we further discuss below, the convergence can be non-monotonic for poorer trial 
wave functions.

\begin{figure}[htbp]
  \includegraphics[scale=0.7]{\PLOTFILE{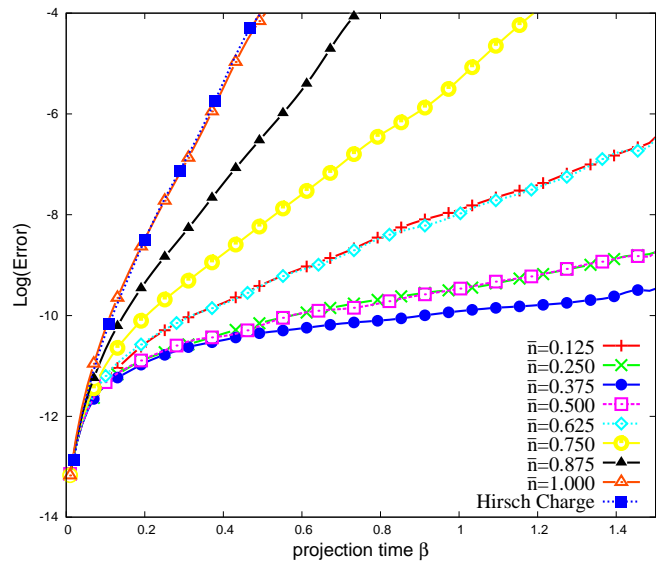}}
  \caption{\label{fig:bg-error}%
(Color online)
Statistical error bar (log-scale) versus projection time for different HS transformations. 
FP calculations are shown. The 
error bars increase exponentially with projection time, but the optimal choice of the 
background $\bar{n}$ in Eq.~(\ref{equ:bg-form}) greatly reduces the fluctuation and improves efficiency. 
The system shown is a $4\times4$ lattice with $N_{\uparrow}=3$ and $N_{\downarrow}=3$, and $U=4$. 
}
\end{figure}

\section{Symmetry Improved AFQMC: Results and Discussion} \label{sec:symmetry}

In this section we describe approaches to impose symmetry in AFQMC calculations,
and study their effects on computational efficiency and, more importantly, on the sign problem.
We divide the discussion into two parts: the choice of the 
Hubbard-Stratonovich transformation and its effect on symmetry, and symmetry in the trial wave function. 
While we will consider all three flavors of AFQMC introduced in Sec.~\ref{sec:AFQMC},
our focus will be on CP and RC, since these are the most general methods which 
will allow calculations to scale to large system sizes. It is shown that the use of symmetry 
can lead to large reductions in statistical and systematic errors, and alleviate the
 sign problem.

\subsection{Hubbard-Stratonovich Transformation}
\label{ssec: symmetry_HS}

For each form of the two-body interaction, there are different ways to decompose the propagator, leading to different forms
of Eq.~(\ref{eq:HS1}).
Decompositions based on Hartree, Fock, and pairing mean-fields are all possible. Even 
within each mean-field framework, the details can affect the final form of the one-body propagator
and the computational efficiency. When a constraint is imposed, the form of the HS transformation
chosen can affect the systematic accuracy given a form of the trial wave function.
For the Hubbard interaction, for example, the most commonly used HS transformation involves discrete  
auxiliary-fields, due to Hirsch \ccite{PhysRevB.28.4059}. 
The spin form of this decomposition is:
\begin{equation}
\label{equ:hirsch-spin}
e^{-\varDelta \tau U n_{i\uparrow}n_{i\downarrow}}=e^{-\varDelta \tau U (n_{i\uparrow}+n_{i\downarrow})/2}
\sum_{x_{i}=\pm1}\frac{1}{2} e^{ \gamma x_{i} (n_{i\uparrow}-n_{i\downarrow})}\,,
\end{equation}
which results in an Ising-like auxiliary-field for each lattice site. The constant $\gamma$ is determined by
\begin{equation}
\cosh(\gamma)=e^{\varDelta \tau U/2}\,.
\label{eq:HS_posU_gamma}
\end{equation}
It can also be mapped to a charge density form:
\begin{equation}
e^{-\varDelta \tau U n_{i\uparrow}n_{i\downarrow}}=e^{-\varDelta \tau U (n_{i\uparrow}+n_{i\downarrow}-1)/2}
\sum_{x_{i}=\pm1}\frac{1}{2} e^{ \gamma x_{i} (n_{i\uparrow}+n_{i\downarrow}-1)}\,,
\label{eq:HS_negU_discrete}
\end{equation}
with
\begin{equation}
\cosh(\gamma)=e^{-\varDelta \tau U/2}\,.
\end{equation}

A more general HS transformation \ccite{PhysRevLett.3.77},
\begin{equation}
\label{equ:con-hubb}
e^{\hat{A}^2}=\frac{1}{\sqrt{2\pi}} \int_{-\infty}^{\infty} e^{-x^2/2+\sqrt{2}x \hat{A}}dx,
\end{equation}
applies to any two-body operators written in the form of a square. To apply this to the
Hubbard interaction, we write: 
\begin{equation}
\label{equ:bg-form}
\begin{split}
n_{i\uparrow}n_{i\downarrow}=\frac{1}{2} [(n_{i\uparrow}+n_{i\downarrow}-\bar{n})^2-\bar{n}^2 \\
-(1-2\bar{n})(n_{i\uparrow}+n_{i\downarrow})
]\,, 
\end{split}
\end{equation}
where $\bar{n}$ can take any value (including acquiring a dependence on $i$). 
We then let $\hat A = \sqrt{-\Delta\tau U/2}\,(n_{i\uparrow}+n_{i\downarrow}-\bar{n})$ and 
use Eq.~(\ref{equ:con-hubb}) to obtain an HS transformation with continuous fields.  
The constant $\bar{n}$ can be thought of, physically, as
a background term that one subtracts from the one-body operator \emph{prior to} applying 
the HS transformation. This has been pointed out before for Hubbard interactions \cite{PhysRevA.72.053610} 
and for Coulomb interactions \cite{baer:6219,al-saidi:224101}. As discussed below, the optimal choice for 
$\bar{n}$ is to remove all background interactions from the mean-field, by minimizing the 
quadratic first term on the right-hand side in Eq.~(\ref{equ:bg-form}). 
Over  typically densities, this choice leads to significant improvement over other choices including the 
standard Hirsch discrete decomposition \ccite{PhysRevB.28.4059}.

\begin{figure}[htbp]
  \includegraphics[scale=0.7]{\PLOTFILE{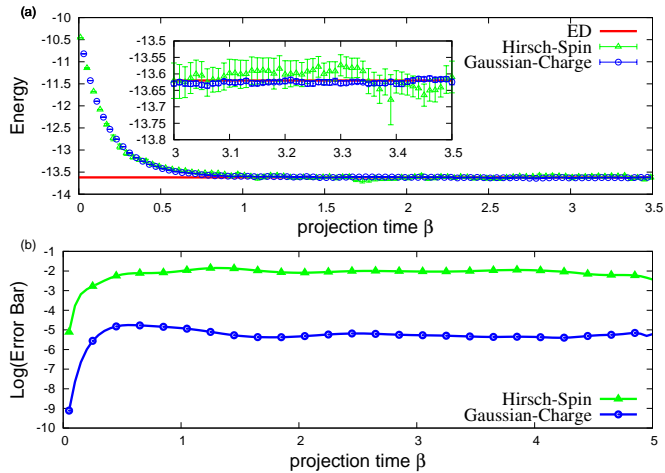}}
  \caption{\label{fig:charge-spin-4488}%
(Color online) Illustration of the effect of preserving symmetry in the HS transformation:
Hirsch spin (Eq.~(\ref{equ:hirsch-spin})) vs.~Gaussian charge (Eqs.~(\ref{equ:con-hubb}) and (\ref{equ:bg-form})). Panel
(a) plots the 
energy from FP versus projection time, with the inset showing a magnified view of $\beta\in(3,3.5)$.
Panel (b) shows the statistical error bar as a function of projection time in a semi-log plot.
The system is the
$8\times8$ Hubbard model, with $N_{\uparrow}=N_{\downarrow}=32$ and $U=8$.
A Hartree-Fock trial wave function is chosen. The number of walkers was 
$10^5$, with a total 20 separate runs to obtain the final averages and estimate the error bars.
}
\end{figure}

In Fig.~\ref{fig:bg-error}, we illustrate the effect of the background $\bar{n}$ in the continuous
charge decomposition of Eqs.~(\ref{equ:con-hubb}) and (\ref{equ:bg-form}).
The  logarithm of the statistical error bar is plotted 
versus projection time for different values of the background $\bar{n}$. 
The calculations are all FP so the sign problem is present, as indicated by the growing error bars,
which are essentially linear in the log-plot with projection time. 
It is seen that the minimum statistical error is achieved when 
$\bar{n}=\langle n_{i\uparrow}+n_{i\downarrow}\rangle_{\rm MF}=(N_{\uparrow}+N_{\downarrow})/L$. 
The efficiency of the HS decomposition decreases as $\bar{n}$ deviates from the optimal 
value. It is a symmetric function of the deviation: a background value which is larger or smaller 
than the optimal value by the same amount gives comparable results. We point out that, although 
we have illustrated this with the repulsive model, the same applies to the attractive case. For example,
in dilute Fermi gas simulations, Eqs.~(\ref{equ:bg-form}) with a small $\bar{n}$ will be much more 
efficient than Eqs.~(\ref{eq:HS_negU_discrete}) which corresponds to $\bar{n}=1$ .

It is often thought that the 
use of a discrete HS field, compared to continuous fields, leads to significant performance
advantages~\cite{PhysRevB.33.3519}. This is not the case, as shown in Fig.~\ref{fig:bg-error}. 
The discrete charge decomposition of Eq.~(\ref{eq:HS_negU_discrete}) 
is shown in the figure. 
We see that 
it is almost the same as the continuous decomposition with
 $\bar{n}=1$. This is because the interaction term in the discrete
charge decomposition is mapped to 
 $(n_{i\uparrow}+n_{i\downarrow}-1)$, identical to the continuous transformation when $\bar{n}$ is set to $1$.
The discrete decomposition is ideal near half-filling, but will be inefficient
in dilute systems, for example, in Fermi gas simulations \ccite{PhysRevA.84.061602}.
  
\begin{figure}[htbp]
  \includegraphics[scale=0.7]{\PLOTFILE{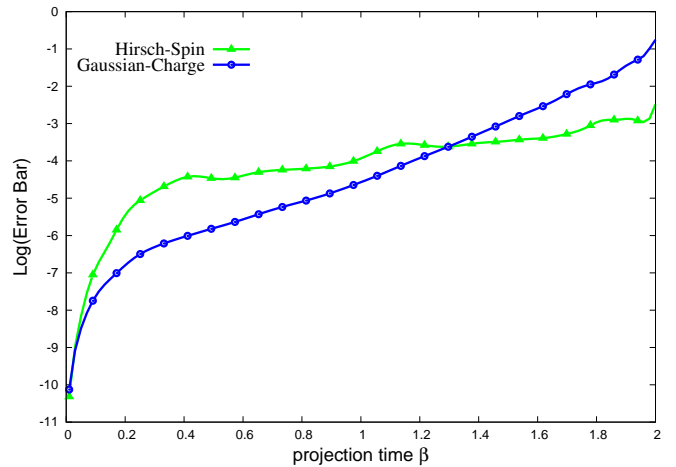}}
  \caption{\label{fig:charge-spin-4477-error}%
(Color online) Comparison of discrete spin and Gaussian charge decomposition in the presence of a sign/phase
problem.
The logarithm of the statistical  error bars from FP is plotted vs.~projection time for a 
$4\times4$ lattice with $N_{\uparrow}=N_{\downarrow}=7$ and $U=8$. Symmetry multi-determinant
 trial wave function is used. The number of walkers is $5\times 10^5$, with total 100 separate runs to estimate the error bars.}
\end{figure}

Because the decomposition in  Eq.~(\ref{equ:con-hubb}) preserves SU(2) symmetry, it 
can be more efficient than the discrete spin decomposition of Eq.~(\ref{equ:hirsch-spin}), which is the most commonly 
used form in simulations of systems with repulsive interactions. This point is more 
subtle, however, as it is intertwined with the sign/phase problem. 
In Fig.~ \ref{fig:charge-spin-4488}, the two decompositions are compared in a situation free
of the sign problem, namely the half-filled repulsive Hubbard model. It can be seen that 
the Gaussian charge decomposition leads to much smaller statistical fluctuations. 
This has also been pointed out by Meng {\it et.~al.}\ccite{Assadnature}, 
using the discrete charge decomposition of Eq.~(\ref{eq:HS_negU_discrete}) at half filling.

In Fig.~\ref{fig:charge-spin-4477-error} we study the case when a phase problem is present (U$>$0, so $\hat{A}$ is imaginary):
$4\times4$, with $N_{\uparrow}=N_{\downarrow}=7$ and $U=8$. 
This system has a severe sign problem for the discrete spin decomposition. 
The continuous Gaussian decomposition leads to a phase problem. 
The latter decomposition initially has smaller error bars, benefiting from the preservation of symmetry, 
but after some time, its error bars exceed that of the spin 
decomposition. So in systems with a sign/phase problem, the efficiency is a balance of two competing aspects. 
On the one hand, the charge decomposition has an advantage for 
preserving symmetry. On the other hand, the phase problem tends to result in fast 
deterioration of the statistical signal and is a disadvantage. Below we discuss how 
to exploit these characteristics in different calculations. 

\begin{figure}[htbp]
  \includegraphics[scale=0.7]{\PLOTFILE{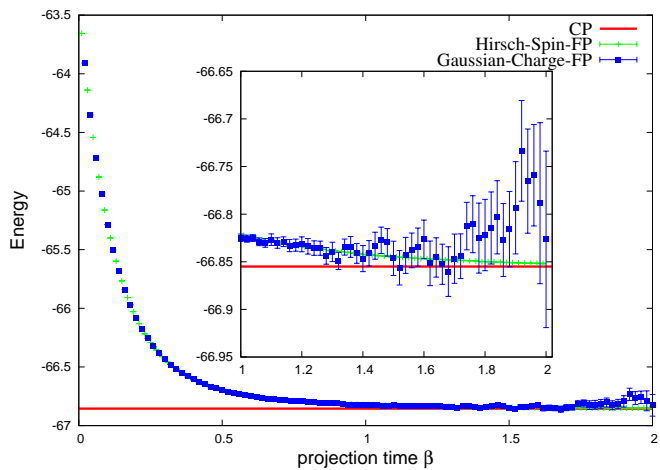}}
  \caption{\label{fig:charge-spin-881313}%
 (Color online) Total energy versus  projection time in 
 FP calculations using discrete spin and continuous charge decompositions.
 The system is a 
  $8\times8$ lattice under periodic boundary conditions, with
 $ N_{\uparrow}=N_{\downarrow}=13$ and $U=4$. A FE
 trial wave function is used. The number of walkers is $5\times10^5$, with total 60 runs to obtain the average and estimate the statistical error bars. The red horizontal line gives the final CPMC result.
 The inset shows a magnified view of the last part of the projection. The spin decomposition 
 calculation provides a reliable upper bound to the ground-state energy.}
\end{figure}

We can use the advantage of the spin decomposition in longer time projections, as shown in Fig.~\ref{fig:charge-spin-881313} in a FP calculation 
 of the periodic $8\times8$ supercell with $N_{\uparrow}=N_{\downarrow}=13$ and $U=4$. 
 The continuous charge decomposition has much larger noise 
 and is not accurate enough. The discrete spin decomposition has rather small fluctuations and 
 provides a useful estimate of an upper bound of the ground-state energy:
 $-66.855(2)$. The run took $\sim17$ hours on 100 AMD Opteron 2.4GHz cores. 
 (For comparison, the CP 
 calculation using a free-electron trial wave function gives $-66.857(2)$, as indicated by the 
 line in Fig.~\ref{fig:charge-spin-881313}, running for 
  minutes on a single core. )
 
\begin{figure}[htbp]
  \includegraphics[scale=0.7]{\PLOTFILE{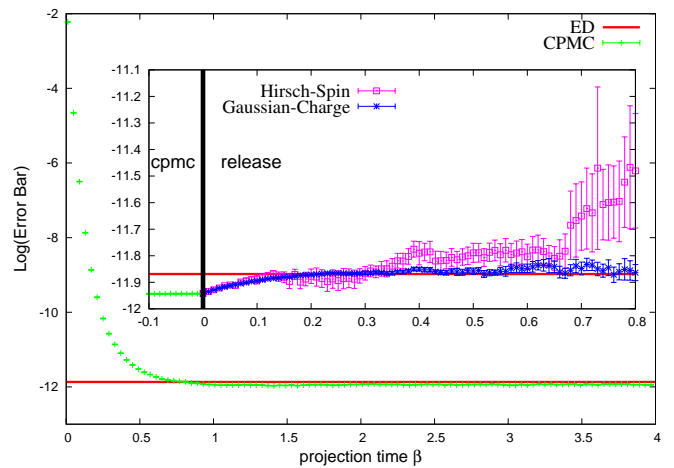}}
  \caption{\label{fig:charge-spin-4477-rcpmc}%
 (Color online) Effect of symmetry decomposition in release constraint calculations. 
 The main figure shows the convergence of
 CPMC energy with projection time, compared with the exact ground-state energy.
  The CPMC calculation uses the discrete spin decomposition.
 The inset shows RC calculations, starting from a converged CP state, using two different 
 forms of the HS decomposition. The system is the same as in Fig.~\ref{fig:charge-spin-4477-error}:
 $4\times4$, $N_{\uparrow}=N_{\downarrow}=7$ and $U=8$, using a symmetry multi-determinant
 trial wave function. The number of walkers is $1\times 10^5$, with total 100 runs in the RC portion 
 to collect statistics. } 
\end{figure}

The charge decomposition can offer a significant advantage in short projection time, however. 
A main application of this is in RC calculations. Since we start from a
population of a converged CP run, the initial state is close to the true ground state. 
One can expect a short projection in the RC calculation to recover
a significant fraction of the correction to CP. In Fig.~\ref{fig:charge-spin-4477-rcpmc},
we show an example RC calculation, for the same system as in Fig.~\ref{fig:charge-spin-4477-error}.
In this calculation, the CPMC portion always 
uses the standard spin decomposition, which has a severe  sign problem. 
As can be seen from the inset,
the CP energy obtained from the mixed-estimate is not variational \ccite{PhysRevB.59.12788}. 
(Typical CP calculations will run to much larger $\beta$ than shown in the main figure, in order to collect statistics.)
In the RC portion, two 
different calculations are shown, one continuing to use the discrete spin decomposition
while the other switching to the symmetry charge-decomposition. It is seen that the latter leads 
to much smaller statistical fluctuations and allows the RC calculation to reach convergence. 
The RC with discrete spin decomposition has much larger errors, and also displays a 
population control bias \cite{PhysRevB.55.7464}. (The statistical error and the bias could, of course, be reduced by increasing the population size further.)

\subsection{Symmetry of the Trial Wave Function}
\label{ssec:symmetry_twf}

In this section we discuss the other aspect of symmetry in AFQMC calculations, 
 the use of trial wave functions which preserve symmetry. 
 We generate the trial wave function with particular symmetries: total spin $S^2$,
total momentum $\overrightarrow{K}$, rotational symmetry in momentum space $R$,
mirror reflection $\sigma$ along the line $K_y=K_x$ in momentum space. When the total momentum $\overrightarrow{K}=0$,
we use the $C_{4v}$ point group irreducible representation to label different symmetry state.
In the present paper, these properties are imposed in the trial wave function 
by a brute-force approach, making
a linear combination of Slater determinants, as discussed in further details below.

\subsubsection{Trial Wave Function in FP Calculations}

Imposing the proper symmetry in the trial wave function can accelerate convergence and reduce 
the equilibration time in the FP calculations. As mentioned in Sec.~\ref{ssec: method_FP}, the trial wave function is often also used to generate the initial population in FP. In all the FP calculations in this section, 
we use the HS transformation given in Eqs.~(\ref{equ:con-hubb}) and 
(\ref{equ:bg-form}), using optimal background values from simple mean-field calculations. 

\begin{figure}[htbp]
  \includegraphics[scale=0.7]{\PLOTFILE{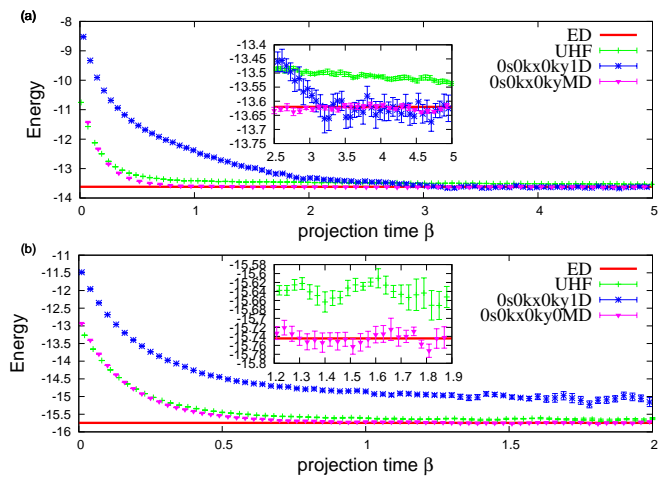}}
  \caption{\label{fig:symmetry-fp}%
  (Color online) The effect of symmetry trial wave functions in FP calculations, compared to exact results in 
   $4\times4$ lattices with $U=4$:
(a) $N_{\uparrow}=N_{\downarrow}=8$, no sign problem; 
(b) $N_{\uparrow}=N_{\downarrow}=7$, severe phase problem.
The UHF trial wave function is generated by a UHF calculation with  $U=0.5$, 
 $0s0kx0ky1D$ is a single-determinant non-interacting trial wave function with
 $S^2=0$ and $k_x=k_y=0$, and 
 $0s0kx0kyMD$ is a multi-determinant trial wave function which has rotational symmetry 
in momentum space in (a) and B1 symmetry in (b), in addition to $S^2=0,k_x=0,k_y=0$.
}
\end{figure}

We illustrate the effect of the symmetry in the trial wave function in Fig.~\ref{fig:symmetry-fp}.
In the top panel, we consider a half-filled system which is thus sign-problem-free. 
The effects of three trial wave functions are compared: the unrestricted Hartree-Fock (UHF) wave function,
a single-determinant trial wave function formed by occupying $k$-states, and a multi-determinant trial wave function which preserves additional symmetry. The UHF trial wave function builds in 
correlation effect via its static anti-ferromagnetic order, and is an excellent starting point, as 
can be seen from the variational energy values at the beginning of the projection. However, it 
is contaminated by higher spin eigenstates, and the FP with UHF exhibits 
a long convergence time, as seen in the inset. (The effect of spin-contamination in AFQMC 
calculations has been discussed in continuum systems by Purwanto {\it et.~al.\/}\ccite{purwanto:114309}.) The single-determinant FE trial wave function has 
a very high variational energy. Its statistical error bars are larger since the $|\psi_{T}\rangle$ in 
Eq.~(\ref{eqn:fpmear}) to evaluate the energy is much poorer, and preserves fewer symmetry properties.
However, it eventually leads to a faster convergence than the UHF trial wave function because symmetry has
properly removed certain excitations. The multi-determinant trial wave function with symmetry,
which we obtain by diagonalizing in the subspace of the open-shell [in the spirit of a small 
complete active-space self-consistent field (CASSCF) calculation], leads to rapid convergence and small 
statistical errors.
In the bottom panel, Fig.~\ref{fig:symmetry-fp}b, we show an example when there is a severe sign/phase problem. The same trends are seen, with the full symmetry trial wave function leading to 
rapid convergence of the projection.

\begin{figure}[htbp]
  \includegraphics[scale=0.7]{\PLOTFILE{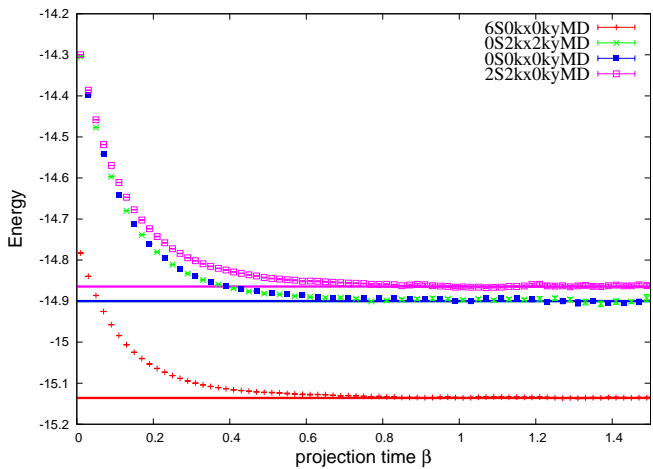}}
  \caption{\label{fig:symmetry-excit}%
(Color online) FP calculations for the ground state and three excited states.
The energy is plotted versus projection time for a $4\times4$ lattice with $N_{\uparrow}=N_{\downarrow}=3$ and $U=4$. The exact results for the ground state and the three excited state energies are shown for comparison.
The symmetry of each energy level is labeled. 
The trial wave functions are chosen with the correct symmetry using multi-determinant.  
The error bars are shown but are 
smaller than symbol size at most points.
}
\end{figure}

With the mixed estimate, the symmetry
trial wave function on the left will have zero overlap with any wave function component in a different symmetry space. 
This allows one to project out explicitly lower energy states of different symmetry, and thus an 
opportunity to study excited states. This has been used in QMC calculations 
before. In the AFQMC formalism, the walkers
 are full Slater determinants, so that the symmetry projection can be done rigorously and 
explicitly for each walker. 
We illustrate excited state calculations in Fig.~\ref{fig:symmetry-excit}, where the converged FP results 
show excellent agreement with results from exact diagonalization. 

\subsubsection{Trial Wave Function in CP Calculations}

In CPMC, the sign or phase problem is controlled by the sign or gauge condition 
of the overlap with the trial wave function. The condition is approximate, and the resulting 
systematic error depends on the trial wave function. Thus the trial wave function 
has an especially important role in CP calculations. In this section, we study how  
trial wave functions which preserve symmetries impact the accuracy in CP calculations.
As shown in prior studies  \ccite{PhysRevB.55.7464,PhysRevB.78.165101}, the 
systematic error from the constraint is in general small for Hubbard-like systems,
even when an FE or UHF trial wave function is used. In closed-shell systems in 
particular, the error is often negligible, as seen in the example of 
$8\times 8$ system with $26$ electrons in Sec.~\ref{ssec: symmetry_HS}. 
The systematic errors tend to be larger for open-shell systems. As we show below,
the leading reason for the problem in open-shell systems seems to be symmetry in the trial 
wave function. The use of trial wave functions with proper symmetry often leads to a dramatic  
reduction in the CP error.
 
\begin{figure}[htbp]
  \includegraphics[scale=0.7]{\PLOTFILE{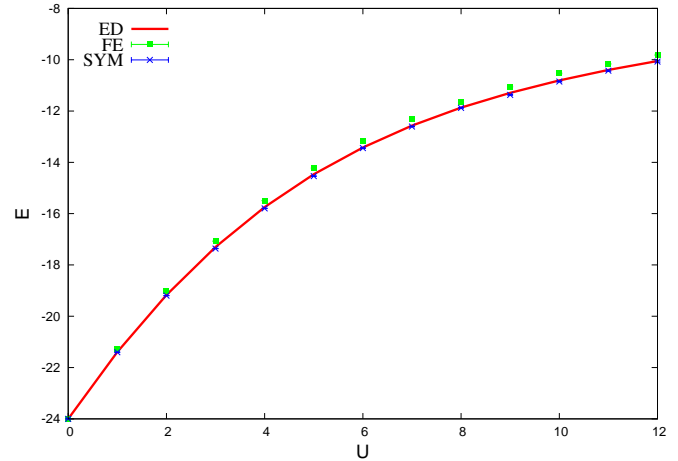}}
  \caption{\label{fig:cpmc-symm-var-u}%
(Color online) Accuracy of CPMC in 
the Hubbard model  
as a function of interaction strength.
Results are shown for the $4\times4$ lattice with $N_{\uparrow}=N_{\downarrow}=7$ 
as a function of $U$, and compared with exact diagonalization. 
When the trial wave function preserves symmetry, the systematic bias in the calculated energy from the CP 
approximation is reduced.}
\end{figure}

We use the discrete spin decomposition in Eq.~(\ref{equ:hirsch-spin})  in the CP calculations, 
which causes ``only'' a sign problem, even when a twist angle is applied in the boundary condition 
of the supercell \ccite{PhysRevB.78.165101}. We first focus on small system sizes where exact results are available to make detailed
and systematic comparison. Larger systems are treated later, and compared with our best results
from RC calculations. 
In Fig.~\ref{fig:cpmc-symm-var-u}, we study the systematic error in the case of 
$4\times4$ with $N_{\uparrow}=N_{\downarrow}=7$, which has the most severe sign problem in 
systems that can be diagonalized presently. We study the systematic error as 
$U$ goes from $0$ to $12$, spanning weak to moderate to strong interactions. 
The CP results with an FE trial wave function is shown. (We use a small twist of opposite sign for 
$\uparrow$ and $\downarrow$ spins to generate the FE trial wave function, which breaks the SU(2) symmetry, 
but has translational symmetry and $\overrightarrow{K}=0$.)
The CP systematic error tends to grow with $U$, 
reaching about $2$\% of the total energy, or about $1$\% of the correlation energy at $U=12$. 
As we see, the use of symmetry wave functions (obtained by diagonalizing the open-shell, leading 
to a total of 10 determinants) makes the CP systematic error very small across the 
range of $U$. Figure~\ref{fig:charge-spin-4477-rcpmc} contains a zoomed-in view, at $U=8$, of the CP/SYM
run and the subsequent RC which leads to an exact result.
In Table~\ref{tab:cpmctwf},  we compile the results from a variety of systems
where exact diagonalization can be done to provide a quantitative measure. The CP results are compared for FE
(or UHF solution obtained from a weak $U$) trial wave functions and symmetry trial wave functions.
Significant improvement is seen in open-shell systems, and accurate results are obtained 
from CP calculations.

In Table~\ref{tab:cpmctwf}, we have included a set of results for  
$4\times4$ and $N_{\uparrow}=N_{\downarrow}=8$. There is no sign 
problem at half-filling, As has been discussed before \ccite{PhysRevB.55.7464,PhysRevB.78.165101}, the CP calculations can be easily made exact at half-filling 
(or for negative $U$ \ccite{PhysRevA.84.061602}) by re-defining the importance sampling 
to have a non-zero minimum.  However, if this were ignored and the CP algorithm applied 
to half-filling literally, an artificial constraint would result because the random walk cannot tunnel 
from one side of $\langle \Psi_T|\phi\rangle=0$ to the other, even though both sides are positive. 
The calculated energy would then show a bias, which is visible in the results shown in the table.
With symmetry trial wave functions, this bias is removed even when running CP unmodified, and
the CP results at half-filling are accurate.  

\begin{figure}[htbp]
  \includegraphics[scale=0.7]{\PLOTFILE{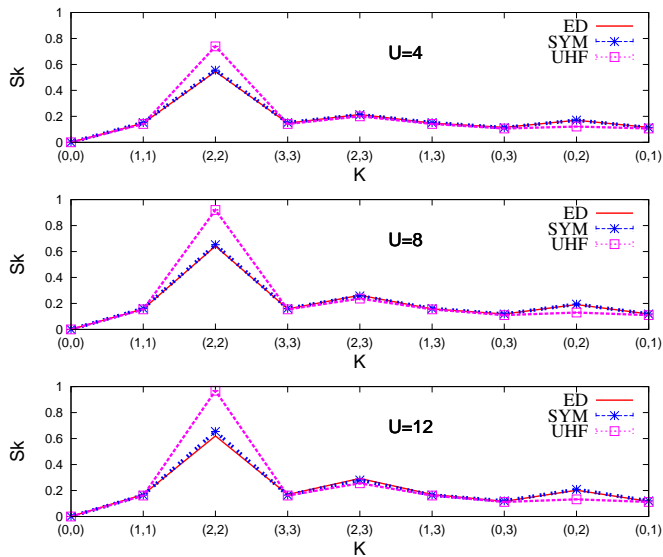}}
  \caption{\label{fig:sk-with-sym} 
  (Color online) The structure factor $S(k)$ of the spin-spin correlation function for three interaction strengths.
  The system is $4\times4$ with $N_{\uparrow}=N_{\downarrow}=7$, and the 
 horizontal axis labels of $K$ are in units of $\pi/2$.
 The symmetry trial wave function has $S^2=0$ and $K_{x}=K_{y}=0$ and B1 symmetry. 
CPMC
has $10,000$ walkers, with back-propagation $\beta=1$. 
}
\end{figure}

The improvement of CP calculations 
with the symmetry trial wave function is not just to the ground-state energy. The CP bias  
in the observable is also significantly reduced. An example is shown in Fig.~\ref{fig:sk-with-sym}, 
in which we calculate the structure factor of the spin-spin correlation function in the ground 
state:
\begin{equation}
 S(K)= 1/N\sum_{ij}S^{z}_{i}S^{z}_{j} \exp[\imath  K ({R_{i}-R_{j}})]\,.
\end{equation}
 As mentioned earlier, we use the back-propagation technique \ccite{PhysRevB.55.7464,PhysRevE.70.056702} 
to calculate correlation functions. 
 The result is plotted 
 for the same $4\times 4$ systems for three different values of $U$. 
  The peak at $(\pi,\pi)$ indicates strong anti-ferromagnetic correlations.
  We see that the CP result using UHF trial wave function shows a larger anti-ferromagnetic order,
  because the UHF state itself over-estimates the order. The symmetry trial wave function 
  removes the bias and leads to results in agreement  with exact diagonalization.

\begin{table*}
\caption{\label{tab:cpmctwf} Computed ground-state energy per site ($E/L$) from CP, with FE
or UHF trial wave function (CP/FE) and with multi-determinant symmetry trial wave function 
(CP/SYM) respectively, compared with release-constraint (RC/SYM)
and exact diagonalization  results. RC/SYM uses 
the same symmetry trial wave function as in CP/SYM. The symmetry of the ground state is given in the 
last column.
 The  statistical error bars in the QMC results are on the last digit and are shown in parentheses.
}
 \begin{ruledtabular}
  \begin{tabular}{ccccccc}
L & ($N_{\uparrow}, N_{\downarrow}, U$)& CP/FE& CP/SYM&RC/SYM& ED &($S^2, K_x, K_y$)\\
\hline
$2\times2$ &(2,1,4)  &-1.60564(5)& -1.60615(3) &-1.60465(5)&-1.60463& (0.75,0,1)\\
$2\times3$ &(2,2,4)  &-1.38328(6) & -1.40129(4)&-1.40085(4)&-1.40087& (2,0,0)\\
$2\times3$ &(2,2,8)  &-1.2239(2)&-1.2463(1)&-1.2443(3)&-1.2442&(2,0,0) \\
$2\times4$ &(2,2,4)  &-1.36839(2)& -1.37387(2)&-1.37379(3)&-1.37383&(0,0,2)\\
$2\times4$ &(3,3,4) &-1.56939(4)&-1.56942(4)&-1.56944(5)&-1.56941&(0,0,0)\\
$3\times3$ &(4,4,8)  &-0.7783(2)& -0.8127(1) &-0.8091(1)&-0.8094&(0,0,0)\\
$4\times4$ &(2,2,4) &-0.72026(5)&-0.72094(1)&-0.72063(1)&-0.72064&(0,0,0) \\
$4\times4$ &(2,2,8) &-0.7070(1)&-0.7082(1)&-0.7075(2)&-0.7076&(0,0,0) \\
$4\times4$ &(2,2,12) &-0.6997(1)&-0.7010(1)&-0.7002(3)&-0.7003&(0,0,0) \\
$4\times4$ &(3,3,4) &-0.93394(2)&-0.94622(1)&-0.94598(1)&-0.94600 &(6,0,0) \\
$4\times4$ &(3,3,8)  &-0.9034(1)&-0.9208(1)&-0.9203(1)&-0.9202&(6,0,0) \\
$4\times4$ &(3,3,12)  &-0.8867(1)&-0.9067(1)&-0.9062(3)&-0.9061&(6,0,0) \\
$4\times4$ &(4,4,4) &-1.09442(2)&-1.09693(2)&-1.09597(6)&-1.09593&(0,0,0)  \\
$4\times4$ &(4,4,8)  &-1.0265(1)&-1.0307(1)&-1.0282(5)&-1.0288&(0,0,0)  \\
$4\times4$ &(4,4,12) &-0.9914(1)&-0.9962(1)&-0.9940(3)&-0.9941&(0,0,0)  \\
$4\times4$ &(5,5,4) &-1.22368(2)&-1.22368(2)&-1.22380(4)&-1.22381&(0,0,0)  \\
$4\times4$ &(5,5,8) &-1.0948(1)&-1.0948(1)&-1.0942(2)&-1.0944 &(0,0,0)  \\
$4\times4$ &(5,5,12) &-1.0292(1)&-1.0292(1)&-1.0278(4)&-1.0284&(0,0,0)  \\
~$4\times4$ &(6,6,4) &-1.1012(1)&-1.1104(1)\footnotemark[1]&-1.1084(2)&-1.1080&(0,0,0)  \\
~$4\times4$ &(6,6,8) &-0.9293(1)&-0.9376(1)\footnotemark[2]&-0.9329(5)&-0.9328&(0,0,0)  \\
~$4\times4$ &(6,6,12) &-0.8439(1)&-0.8557(1)\footnotemark[2]&-0.8507(6)&-0.8512&(0,0,0)  \\
~$4\times4$ &(7,7,2) &-1.19584(2)&-1.19992(1)\footnotemark[3]&-1.19822(2)&-1.19821&(0,2,2) \\
~$4\times4$ &(7,7,4) &-0.9793(1)&-0.9863(1)\footnotemark[1]&-0.9840(1)&-0.9840&(0,0,0) \\
~$4\times4$ &(7,7,6) &-0.8334(1)&-0.8428(1)\footnotemark[1]&-0.8386(3)&-0.8388&(0,0,0) \\
~$4\times4$ &(7,7,8) &-0.7361(1)&-0.7461(1)\footnotemark[1]&-0.7417(8)&-0.7418&(0,0,0) \\
~$4\times4$ &(7,7,10)&-0.6687(2)&-0.6782(1)\footnotemark[1]&-0.673(2)&-0.6754&(0,0,0) \\
~$4\times4$ &(7,7,12)  &-0.6202(2)& -0.6296(2)\footnotemark[1]&-0.627(4)& -0.6282 &(0,0,0) \\
$4\times4$ &(8,8,4)&-0.84225(6)&-0.85140(6)&-0.85133(6)&-0.85137&(0,0,0) \\
$4\times4$ &(8,8,8)&-0.5164(2)&-0.5293(2)&-0.5291(2)&0.5293&(0,0,0) \\
$4\times4$ &(8,8,12)&-0.364(3)&-0.3741(2)&-0.3739(4)&-0.3745&(0,0,0)
  \end{tabular}
 \end{ruledtabular}	
\footnotetext[1]{$B_1$ symmetry is also used in $|\psi_T\rangle$.}
\footnotetext[2]{$A_1$ symmetry is also used in $|\psi_T\rangle$.}
\footnotetext[3]{$\hat{R}_{\pi/2} |\psi_T>=\exp(i 3 \pi/2) |\psi_T>$ symmetry is also used.
 }
\end{table*}

\subsubsection{Trial Wave Function in RC Calculations}

Formally the role of symmetry in the trial wave function in RC calculations is similar to 
that in FP. However it is intimately connected to the discussion in the previous section on CP, 
since the initial state in RC is the converged solution from CP. The symmetry trial wave functions 
improves the CP approximation and the quality of the
wave function sampled from CP, as indicated by the improvement in the energy and in the 
calculated observables. This means symmetry trial wave functions also allow better 
RC calculations, by providing a better initial state \emph{and} by giving a better trial wave function 
in the mixed estimate in Eq.~(\ref{eq:E_mixed_RC}). As discussed in Sec.~\ref{ssec: symmetry_HS}, 
we also impose symmetry with the HS transformation in RC, by switching from the Ising spin form in the CP 
calculation to the Gaussian form in the RC part. We find this combination to   
improve the quality of the RC calculations greatly in most cases. An example  
is shown in Fig.~\ref{fig:charge-spin-4477-rcpmc}.
Results from RC/SYM calculations are also shown 
in Table~\ref{tab:cpmctwf} 
for systematic comparisons  with
 CP and with exact diagonalization results.

\begin{figure}[htbp]
  \includegraphics[scale=0.7]{\PLOTFILE{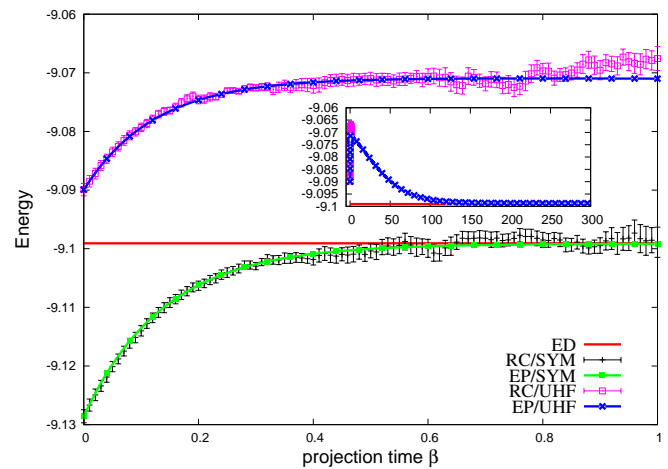}}
  \caption{\label{fig:sym-release} 
  (Color online) RC calculations with symmetry trial wave functions and without. 
  The system is  $3\times 3$ with $N_{\uparrow}=
N_{\downarrow}=2$ and $U=4$. 
The symmetry trial wave function has $S^2=2,K_{x}=0,K_{y}=0$ while the UHF 
wave function breaks these symmetries. CP/UHF is very accurate, but RC/UHF 
has non-monotonic behavior and slow convergence, as shown by the explicit propagation. 
RC/SYM converges rapidly and monotonically. The 
explicit propagation (EP) 
result of RC/UHF is shown to large projection time in the inset. 
}
\end{figure}

Figure~\ref{fig:sym-release} illustrates the behavior of RC calculations using two different 
trial wave functions, the UHF versus a symmetry trial wave function. A small system size of  
$3\times3$ with $2\uparrow$ and $2\downarrow$ electrons is chosen 
such that  the RC calculation can also be carried out explicitly to allow direct comparison. (In the 
explicit calculation, we propagate the CP population of $\{ |\phi_i^{\rm CP}\rangle \}$ directly by
applying $e^{-\Delta \tau \hat H}$. The propagation is carried out by expanding each walker in terms of 
exact eigenstates of $\hat H$.) We see that CPMC/UHF gives an energy closer to the exact value ($\sim 0.1$\% error) compared to
CPMC/SYM ($\sim - 0.3$\% error).  The corresponding RC/UHF moves further away from the exact answer and shows 
no indication of convergence in the imaginary-time span in which RC/SYM is well-converged. 
The explicit RC calculation, as shown in the inset, reveals a highly non-monotonic behavior.
The projection does converge to the correct ground-state energy, but requiring an imaginary time
of $>100$. This would be impossible to reach in a QMC RC calculation because of the 
sign problem. Thus non-monotonic behaviors could be difficult to detect and would yield 
misleading results. The improvement with the symmetry trial wave function, which 
leads to rapid and monotonic convergence, is then especially valuable. 

\begin{figure}[htbp]
  \includegraphics[scale=0.7]{\PLOTFILE{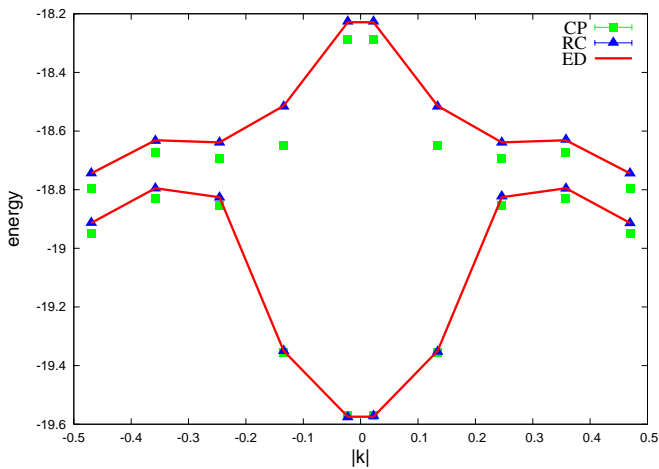}}
  \caption{\label{fig:4-4-5-5-struct}%
(Color online) RC and CP results for ground and the first excited state energies versus 
crystal momentum. 
RC greatly improves the calculation of excited states and band structures. The system is 
$4\times 4$ with $N_{\uparrow}=
N_{\downarrow}=5$ and $U=4$.
QMC statistical error bars are smaller than symbol size.
The horizontal axis gives $|k|$ along a line cut $k_y=2k_x$. Exact diagonalization (ED) results are shown for comparison. The line is to aid the eye. 
}
\end{figure}

The use of proper symmetry can allow RC calculations of excited states, similar to the 
discussion in Fig.~\ref{fig:symmetry-excit}. Since CP allows one to start from an initial 
population much closer to the exact exact state, RC can be more accurate. 
An example is shown in Fig.~\ref{fig:4-4-5-5-struct}, in which the many-body ground state and 
first 
excited state energies are calculated as a function of crystal 
momentum.  Both CP and RC are done with the same trial wave function, in which 
the correct symmetry is imposed.
Consistent with prior experience, CP is very accurate for the ground state,  although systematic error is  visible at larger 
twist angles. The CP result is less accurate for the excited state. With RC, the CP error is removed and the results are seen to be essentially exact. 

\begin{table}
\caption{\label{tab:biglattice} Energy per site in some open-shell
Hubbard square lattices at $U=4$.  The symmetry of the ground state, $S^2=0,K_x=K_y=0$, is preserved in the trial wave function used in CP/SYM and RC/SYM. 
The UHF trial wave function is generated with $U=0.5$ as has been done before 
\cite{PhysRevB.55.7464}.
}
\begin{ruledtabular}
\begin{tabular}{cccccc}
L & ($N_{\uparrow}$ ,$N_{\downarrow}$)&CP/UHF& CP/SYM& RC/SYM\\
\hline
6$\times$6 &(12 ,12)  &-1.18444(3)&-1.18625(3) &-1.18525(4) \\
6$\times$6 &(24 ,24)  &0.14889(2) &0.14709(3)   &0.14809(4)  \\
8$\times$8 &(14 ,14)  &-1.07173(1) &-1.07239(1) \footnotemark[1] &-1.07180(2)\\
8$\times$8 &(22 ,22)  &-1.18580(2)&-1.18673(2)& -1.1858(2)  \\
10$\times$10 &(40,40)& -1.11378(3) &-1.11468(2) & -1.1135(2) \\
12$\times$12 &(58,58) &-1.10912(3) &-1.11015(3) & -1.1089(2)
\end{tabular}
\end{ruledtabular}
\footnotetext[1] {$B_2$ symmetry is also used in $|\psi_T\rangle$.} 
\end{table}

\section{Discussion and Conclusion}\label{sec:conoutl}

In this paper, we have studied the role of symmetry in AFQMC calculations, and discussed 
the imposition of symmetry from two key aspects of an AFQMC calculation, namely the 
HS transformation and the trial wave function. It is shown that major improvements in 
efficiency and accuracy can be achieved. To allow detailed and systematic benchmark and 
analysis, we have used 
smaller lattice sizes extensively, where exact results are available. The method applies 
straightforwardly to larger systems. CP calculations will scale as a low power with system size;
FP and RC will of course have a sign/phase problem, albeit at a much reduced level with the
proposed symmetry improvements.
In Table~\ref{tab:biglattice}, we 
present CPMC energies in several open-shell systems up to $12\times 12$. (For closed shell systems, the single-determinant 
FE trial wave function already satisfies the symmetries. They are expected to be very accurate. We have verified this in several 
cases with RC runs). 
CP/SYM and the corresponding RC results are shown. The RC results in Table~\ref{tab:biglattice} are essentially exact 
ground-state energies.  CP/UHF results are also shown, which are comparable and even closer to the exact answer than the CP/SYM using simple symmetry trial wave functions in these systems. 
(As discussed, the CP/SYM leads to much better convergence in RC calculations.)
This confirms the accuracy of CP/UHF as has been previously asserted. 

We have discussed the general continuous Gaussian charge decomposition, which preserves 
spin symmetry. It is shown that the proper choice of the background $\bar {n}$ can lead to 
large reduction of the statistical fluctuation. 
One advantage of the spin decomposition is 
that, for repulsive interactions, it results in a sign problem in contrast with a phase problem for the charge decomposition. 
CP calculations with the spin decomposition tend to perform much better.
We have emphasized
the idea that the different forms can have advantages in different situations. 
Generally, the merits of the discrete spin and continuous charge, or indeed other
forms of the HS transformation, will depend on the actual problem and physics.
However, preserving the right symmetry is highly valuable, as we have demonstrated. Especially worth 
noting is that the switch to charge-decomposition within a CP calculation of spin-decomposition 
makes much better RC performance over many parameter ranges. 

We have shown the importance of having trial wave functions which preserve symmetry. 
These trial wave functions accelerate convergence, allow better calculations of excited 
states, can significantly reduce the CP systematic error, and make possible 
systematically improvable RC calculations.
The approach we have taken to generate trial wave functions that preserve symmetry,
the equivalent of a small CASSCF calculation in quantum chemistry,
has provided a proof-of-concept. 
In addition, they have already allowed calculations
in significant system sizes at physically important regimes (low doping), as shown in Table~\ref{tab:biglattice}. However, for general open-shell situations, the CASSCF approach
does not scale well. The resulting number of Slater
determinants in the trial wave function will grow rapidly with system size. Several alternatives are possible, including projected 
BCS wave functions \ccite{PhysRevA.84.061602} and projected Hartree-Fock wave functions \ccite{PhysRevB.85.245130}. 

The development presented in this paper will allow many applications even in its current form. 
Although we have 
focused on zero-temperature methods, many of the ideas will also apply to 
finite-temperature calculations \cite{BSS1PhysRevD.24.2278,PhysRevLett.83.2777}. The formula can be directly mapped from the U$>$0 case we discussed to U$<$0 with a particle-hole transformation. Indeed the principle works for any other two-body interactions.
 The 
CP/SYM calculations and the RC/SYM from it represent a major step forward, as we can now have internal checks
and a systematically improvable computational method capable of reaching two- and three-dimensions and large system sizes. 

\begin{acknowledgments}

We thank
S.~Chiesa,
H.~Krakauer,
F.~Ma,
W.~Purwanto,
G.~Scuseria,
and C.~Umrigar
for helpful discussions.
This research was supported by
DOE (Grant No.~DE-SC0008627) and NSF (Grant No.~DMR-1006217).
We also acknowledge a DOE CMCSN award for facilitating stimulating interactions,  
and an INCITE award for computing using resources of the Oak 
Ridge Leadership Computing Facility at the Oak Ridge National Laboratory, which
 is supported by the Office of Science of the U.S. Department of Energy under Contract No.~DE-AC05-00OR22725.

\end{acknowledgments}

\bibliography{afqmc}

\end{document}